\documentclass[structabstract]{aa}  
\usepackage{graphicx}
\usepackage{txfonts}
\usepackage{natbib}
\usepackage{balance}
\bibpunct{(}{)}{;}{a}{}{,} 
\begin{document}
   \title{Historical light curve and search for previous \\outbursts of Nova KT~Eridani (2009)}


   \author{R. Jurdana-{\v S}epi\'{c}\inst{1}
          \and
          V. A. R. M. Ribeiro \inst{2,3,6}\thanks{Correspondence to: vribeiro@ast.uct.ac.za}
          \and
          M. J. Darnley \inst{2}
          \and
          U. Munari \inst{4,5}
          \and
          M. F. Bode \inst{2}
           }

   \institute{Physics Department, University of Rijeka, Omladinska 14, HR-51000 Rijeka, Croatia 
        \and
        Astrophysics Research Institute, Liverpool John Moores University, Egerton Wharf, Birkenhead, CH41 1LD, UK 
        \and
        Astrophysics, Cosmology and Gravity Centre, Department of Astronomy, University of Cape Town, Private Bag X3, Rondebosch 7701, South Africa
        \and
        INAF Astronomical Observatory of Padova, via dell'Osservatorio, 36012 Asiago (VI), Italy 
        \and
	ANS Collaboration, c/o Astronomical Observatory, 36012 Asiago (VI), Italy 
        }

   \date{Received ; accepted}

  \abstract
   {Nova Eridani (2009) caught the eye of the nova community due to its fast decline from maximum, which was initially missed, and its subsequent development in the radio and X-ray wavelengths. This system also exhibits properties similar to those of the much smaller class of recurrent novae; themselves potential progenitors of Type Ia Supernovae.}
   {We aim to determine the nature and physical parameters of the KT~Eri progenitor system.}
   {We searched the Harvard College Observatory archive plates for the progenitor of KT~Eri to determine the nature of the system, particularly the evolutionary stage of the secondary. We used the data obtained to search for any periodic signal and the derived luminosity to estimate a recurrence timescale. Furthermore, by comparing the colours of the quiescent system on a colour-magnitude diagram we may infer the nature of the secondary star.}
   {We identified the progenitor system of KT~Eri and measured a quiescent magnitude of $<B> = 14.7\pm0.4$.  No previous outburst was found. However, we suggest that if the nova is recurrent it should be on a timescale of centuries. We find a periodicity at quiescence of 737 days which may arise from reflection effects and/or eclipses in the central binary. The periodicity and the quiescence magnitude of the system suggest that the secondary star is evolved and likely in, or ascending, the Red Giant Branch. A second period is evident at 376 days which has a sinusoidal like light curve. Furthermore, the outburst amplitude of $\sim9$ magnitudes is inconsistent with those expected for fast classical novae ($\sim17$ magnitudes) which may lend further support for an evolved secondary.}
   {We investigated the probable recurrent nova KT Eri for which we suggest an inter-outburst period of order centuries and an evolved secondary. This may suggest that there is a whole range of possible inter-outburst periods in between the ``typical'' classical and recurrent novae nomenclature. Archival searches are an excellent tool in order to investigate the nature of astrophysical objects, in order to determine the nature and physical parameters.}


   \maketitle
%

\section{Introduction}
\footnotetext[6]{South African Square Kilometre Array Fellow}
A classical nova (CN) outburst occurs due to a thermonuclear runaway on the surface of a white dwarf (WD) in a binary system \citep[see e.g.,][]{BE08,B10}. The companion is typically a Roche lobe filling main-sequence star and the outburst is expected to recur on a timescale of 10$^4$ $-$ 10$^5$ years. A much smaller, and related, sub-group is the recurrent novae (RNe) where outbursts recur on a timescale of decades. These short recurrence times have been attributed to a combination of a massive WD, probably close to the Chandrasekhar limit, and a high mass accretion rate. For example, using typical recurrence timescales for a CN or RN with a 1.25~M$_{\odot}$ WD primary the accretion rates are predicted to be $\sim$10$^{-9}$~M$_{\odot}$ yr$^{-1}$ and $\sim$10$^{-7}$~M$_{\odot}$ yr$^{-1}$, respectively \citep{2005ApJ...623..398Y}.

The RNe can be broadly divided into three main groups according to their outburst properties and the nature of the secondary star; ({\it i}) the RS Oph-type, which constitute a red-giant secondary with orbital periods of order hundreds of days. ({\it ii}) The U Sco-type, with sub-giant secondaries and orbital periods of order days. ({\it iii}) The T Pyx-type, more akin to CN with main sequence secondary stars and orbital periods of order hours \citep[e.g.][]{A08}.

Nova Eridani 2009 (hereafter, KT Eri), located at RA~=~04$^{\mathrm{h}}$47$^{\mathrm{m}}$54.$\!\!^{\mathrm{s}}$21 and Dec = $-$10$^{\circ}$10$^{\prime}$43.$\!\!^{\prime\prime}$1 (J2000), was discovered on 2009 November 25.54 UT in outburst by \citet{I09} at which time a possible progenitor was also identified. The discovery was confirmed by \citet{GSM09}. Early low resolution optical spectra obtained on 2009 November 26.56 showed broad Balmer series, He {\sc i} 5016\AA\ and N {\sc iii} 4640\AA\ emission lines, the FWHM of the H$\alpha$ emission was around 3400 km s$^{-1}$ \citep{M09}. This object was subsequently confirmed as a He/N nova \citep{RPR09}.

KT~Eri was discovered in outburst well after maximum-light.  Other authors later reported the nova clearly in outburst on 2009 November 18 and that the outburst occurred at some point after 2009 November 10.41 \citep{DDG09}.  Using a combination of the Solar Mass Ejection Imager (SMEI) archive and Liverpool Telescope SkyCamT\footnote{http://telescope.livjm.ac.uk/Info/TelInst/Inst/SkyCam} observations \citet{HBH10} confirmed that KT Eri was already in outburst on 2009 November 13.12 and constrained the time of maximum to 2009 November 14.67~$\pm$~0.04 (which we take as $t$ = 0). The band pass for the SMEI is from 4500$-$9500\AA\ with a peak quantum efficiency of the instrument at 7000\AA\ \citep{ESC03}. \citet{HBH10} also reported a significant pre-maximum halt before the luminosity peaked at $m_{\mathrm{SMEI}}$~=~5.42~$\pm$~0.02 and confirmed the very fast nature of the outburst ($t_2=6.6$ days).  \citet{RBS09} used the MMRD relationship with $t_2=8$ days to determine a distance to KT~Eri of 6.5 kpc; the SMEI $t_2$ determination implies a slightly larger distance.

KT~Eri was also detected at radio wavelengths \citep{OMS10} and as a X-ray source \citep{BOP10}.  The radio observations indicate that the nova was observed during the rise of the radio light curve \citep{OMS10}.  The first {\it Swift} XRT detection of KT~Eri was on day 39.8 post-outburst as a hard source \citep[all but one count above 1 keV;][]{BOP10} as was also the case on day 47.5. However, by day 55.4 the super soft source (SSS), usually associated with nuclear burning on the surface of the WD, began to emerge.  On day 65.6 post-outburst the SSS softened dramatically \citep{BOP10}. They also noted that the time scale for emergence of the SSS was very similar to that of the RN LMC~2009a \citep{BOP09}.  Here all times have been corrected to the SMEI determination.

Following the methodology applied to the RN candidates V2672~Oph and V2491~Cyg \citep[][respectively]{MRB11,RDB11}, \citet{R10} modelled the early H$\alpha$ line profile evolution and concluded that the nebular remnant was bipolar, with an inclination of 53$^{+6}_{-7}$ degrees and a maximum expansion velocity of 2800 $\pm$ 200 km s$^{-1}$.  It is noteworthy that this expansion velocity is also comparable to the RN LMC~2009a (Bode et al. in preparation) however, lower than those observed in other RNe (for example, RS~Oph and U~Sco, \citealt{RBD09} and \citealt{MZT99}, respectively) although still higher than a typical CN \citep[e.g.][]{SOD95}.

KT~Eri is of particular interest as it exhibited similar behaviour to known RNe, particularly LMC~2009a (Bode et al., in preparation), in terms of its optical photometric and spectroscopic behaviour and the evolution of its X-ray emission.  In this paper we report the results of a search for the both progenitor system and previous outbursts of KT~Eri, as well as the behaviour of the system at quiescence.  In Section \ref{data} we describe our approach and present the results in Section \ref{results}.  We discuss our findings in Section \ref{discussion} and finally give conclusions in Section~\ref{conclusions}.

\section{Data Summary}\label{data}
\begin{figure*}
\begin{center}
  \includegraphics[width=0.4\textwidth]{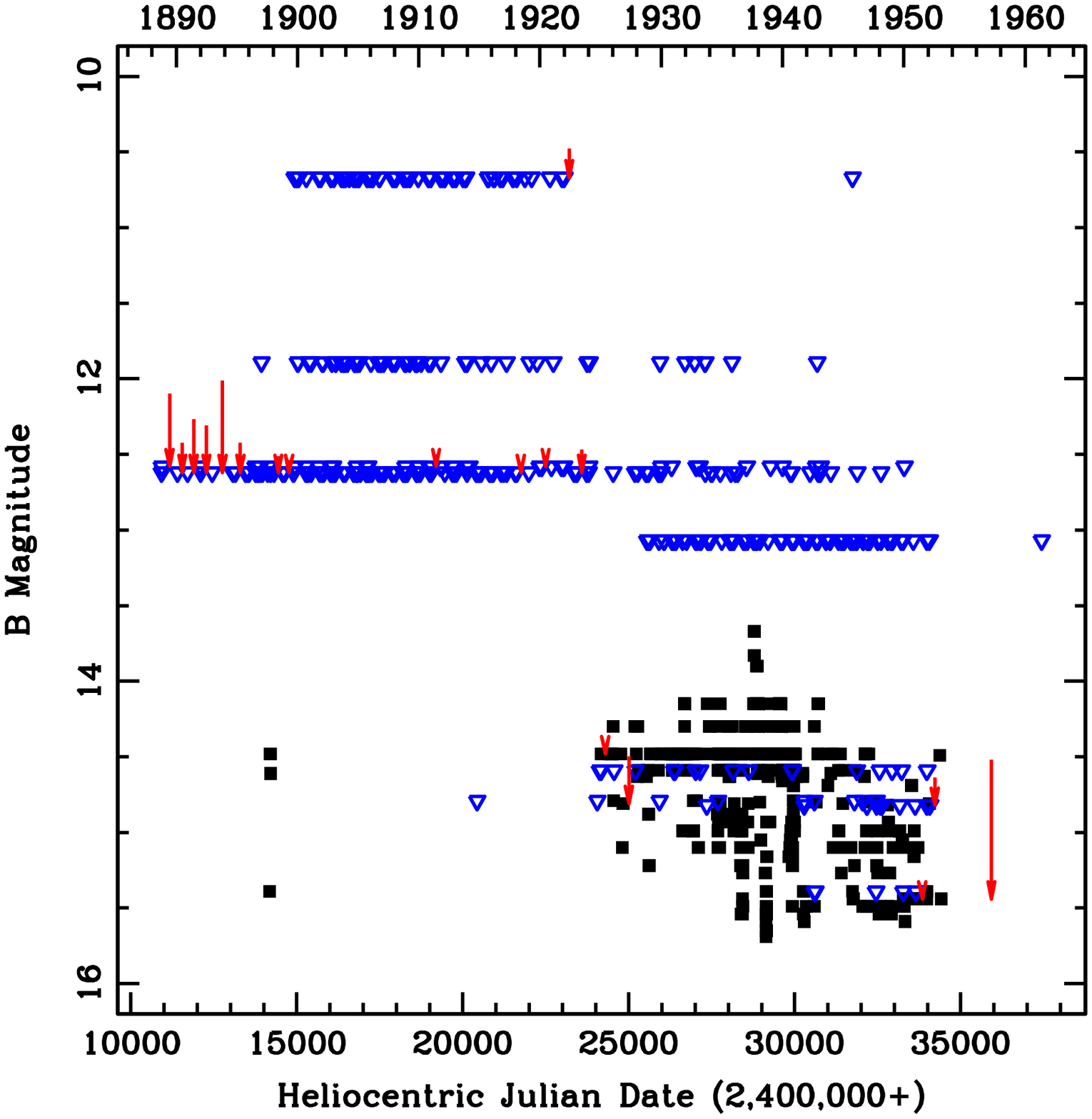}\includegraphics[width=0.4\textwidth]{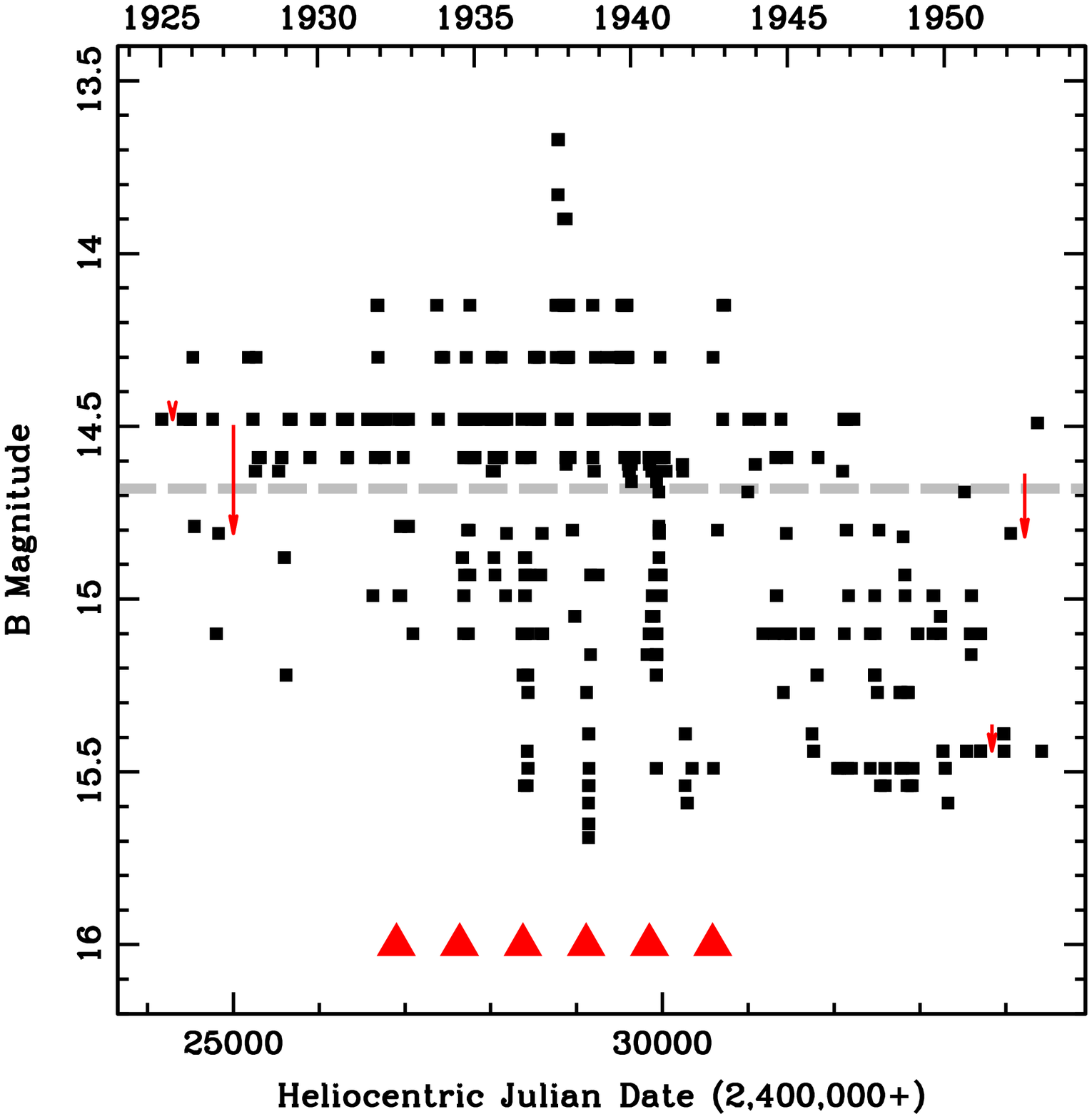}
\end{center}
  \caption{Harvard College Observatory archival light curve of KT~Eri.  The black squares indicate $B$-band magnitudes, the blue triangles upper limits.  The red arrows indicate gaps in the coverage that could accommodate an un-detected outburst similar to that in 2009.  The length of the arrow tails are proportional to the probability of ``missing'' an outburst that occurred between two consecutive Harvard plates (length of one magnitude indicates a 100\% probability). {\bf Left:} Entire light curve between 1885 and 1965.  {\bf Right:} Light curve between 1923.5 and 1954.5.  These data show a highly variable progenitor with $<B>=14.7\pm0.4$ (grey dashed line) and a full amplitude variability of around one magnitude. The red triangles show the derived period of 737 days, which coincides with the dips in the light curve.}
  \label{fig:fig1}
\end{figure*}

The collection of astronomical plates at the Harvard College Observatory contains around half a million observations taken between the mid-1880s and 1989 (with a gap between 1953 and 1968).  The majority of these are blue plates with a limiting magnitude of 15$^{th}$ or brighter. However, some plates are as deep as 18$^{th}$ magnitude.  We obtained 1,012 plates dating from 1888 to 1962 (see Figure~\ref{fig:fig1}). The subsequent analysis of these plates yielded 495 detections and 517 upper limits at the coordinates of KT~Eri.

Plates potentially covering the KT~Eri field were selected for the Harvard general database by virtue of their plate centres, angular extension and orientation on the sky.  The plates were manually retrieved from the plate stack, placed under a high quality monocular lens and the area encompassing KT~Eri and the surrounding comparison stars centred on the eyepiece.  The magnitude of KT~Eri was then visually estimated against the comparison sequence. This method has been used extensively and successfully tested to reconstruct the photometric history of symbiotic stars and novae from several thousand plates in the Asiago plate archive \citep[e.g.][]{MJM01,MJ02,JM10}.  For KT~Eri, we used the same UBVR$_{\textrm{c}}$I$_{\textrm{c}}$ comparison sequence, calibrated against the Landolt (1983, 1992) standard stars, as used in the extensive BVRI CCD photometric monitoring of the 2009 outburst performed by the ANS Collaboration \citep{MSF11}, discussed in a forthcoming paper. This ensures a common magnitude scale between pre- and post-outburst photometric data.

For plates where the KT~Eri progenitor is above the plate limit, a magnitude was recorded and an error estimated.  The latter was derived following comparison stars closest in brightness to KT~Eri (those immediately brighter and fainter). These were carefully inspected on each plate.  If $d$ was the tabulated difference in magnitude, and $n$ the number of steps in which the difference $d$ could still be divided and yet confidently recognised by eye-inspection, the error was then estimated as $err = d / n$ and recorded, rounded to the nearest 0.05 mag.  In this way the error automatically includes the effects due to plate focusing, background fogging, seeing and guiding.  The error does not include, however, the effect of departure from the actual $B$-band transmission of the combination of emulsion sensitivity and filter, optics and atmosphere transmission. The departure is generally pretty well compensated by using a variety of comparison stars recorded on the same plate as the KT Eri progenitor. However, differences in effective temperatures among them reflects into corresponding differences in the effective wavelengths of observations, and this explains a fractional part of the noise in the data. Plates where the progenitor was below the detection threshold, we recorded as an upper limit to the brightness of the nova the value of the faintest of the visible comparison stars.

The Two Micron All Sky Survey \citep[2MASS;][]{SCS06}, United States Naval Observatory B1.0 Catalog \citep[USNO-B1;][]{MLC03} and the Deep Near Infrared Survey of the Southern Sky (DENIS 3rd release) were all employed to enable a search for the progenitor of KT Eri through the NASA Infrared Processing and Analysis Center (IPAC) Infrared Science Archive (IRSA).

\section{Results}\label{results}

\subsection{Search for any previous outbursts}

Using the processed Harvard data, we searched for any previous un-recorded outbursts of KT~Eri. However, as is clearly evident from Figure~\ref{fig:fig1}, there were no additional outbursts contained within the Harvard data in the time span of the available plates. Based on the plate photometry, we identified any gaps in the coverage during which any outburst would have been missed (these are indicated by the red arrows in Figure~\ref{fig:fig1}).  To further investigate the likelihood of any ``missed'' outbursts, we employed a Monte Carlo technique, generating ten million randomly occurring outbursts over the course of the Harvard data span.  The light curve of each seeded outburst was based upon the AAVSO\footnote{see http//www.aavso.org} and SMEI \citep{HBH10} observations of the 2009 outburst.  By taking into account the sensitivity of the plate data at each Harvard epoch we computed the fraction of seeded outbursts occurring between any consecutive pair of Harvard observations that would have been missed despite the Harvard data.  This ``non-detection'' probability is indicated by the length of the red arrow tails in Figure~\ref{fig:fig1}, where an arrow tail length of 1 mag indicates a 100\% probability that all the seeded outbursts would have been un-detected.

\begin{figure}
\begin{center}
\includegraphics[width=0.8\columnwidth]{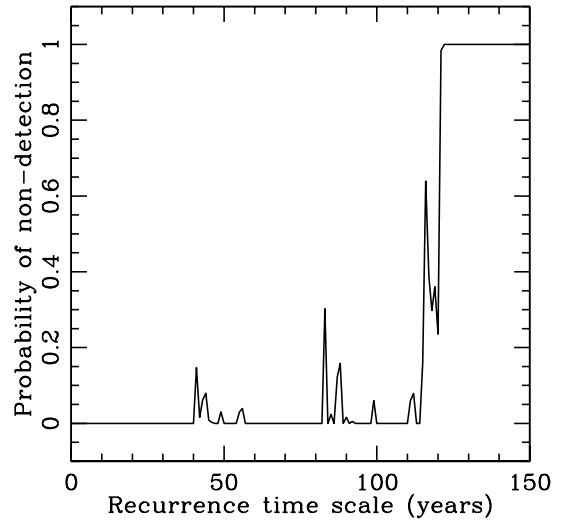}
\end{center}
\caption{A plot showing the probability of all outbursts being un-detected by the Harvard plate archives as a function of the recurrence time scale of KT~Eri.  For example, for a recurrence time scale of 10 years at least one outburst would have been visible in the Harvard data. However, for recurrence times $>$ 120.5 years all outbursts would go undetected as this is the limit of the data span.  If the recurrence time of the system is 41 years then the probability that no outburst was caught by the Harvard data is only 15\%.}
\label{fig:stats}
\end{figure}

In addition, we used the Monte Carlo simulation to compute the fraction of ``missed'' outbursts as a function of recurrence time scale (see Figure~\ref{fig:stats}). The Monte Carlo simulation also seeded outbursts with a given recurrence time scale, between 1 and 200 years in 0.01 year steps (later binned up to complete years for analysis).  Here we assumed that all outbursts are identical and that the intra-outburst period is constant \citep[for example, U Sco --][for which this is approximately true]{S10}.  The data presented in Figure~\ref{fig:stats} shows that for a wide range of recurrence time the probability of an outburst not appearing within the Harvard data is zero.  This probability jumps to 100\% for recurrence times $>120.5$ years as this is the limit of the Harvard data.  However, there are small clusters of recurrence times that cannot be completely ruled out, despite the Harvard data.  These recurrence timescales are $40.5-44.5$ years and multiples there of.  For example, given a recurrence time of 41 years, the probability that all three outbursts expected to appear within the Harvard data were missed, despite the Harvard data, is only 15\%.

\subsection{Periodicity}

We also found that the source is variable by $\sim1$ magnitude at quiescence, as shown in Figure~\ref{fig:fig1}. This variability is greater than the $3\sigma$ error on the plate photometry, and the mean quiescent magnitude is $B = 14.7\pm0.4$~mag (quoted error is one standard deviation).  The light curves of quiescent novae typically show two forms of variation; essentially random fluctuations due to flickering of any accretion disk, and periodic variations usually related to the orbital period of the system such as eclipses or reflection effects.

\begin{figure*}
\begin{center}
  \includegraphics[angle=270,width=0.8\textwidth]{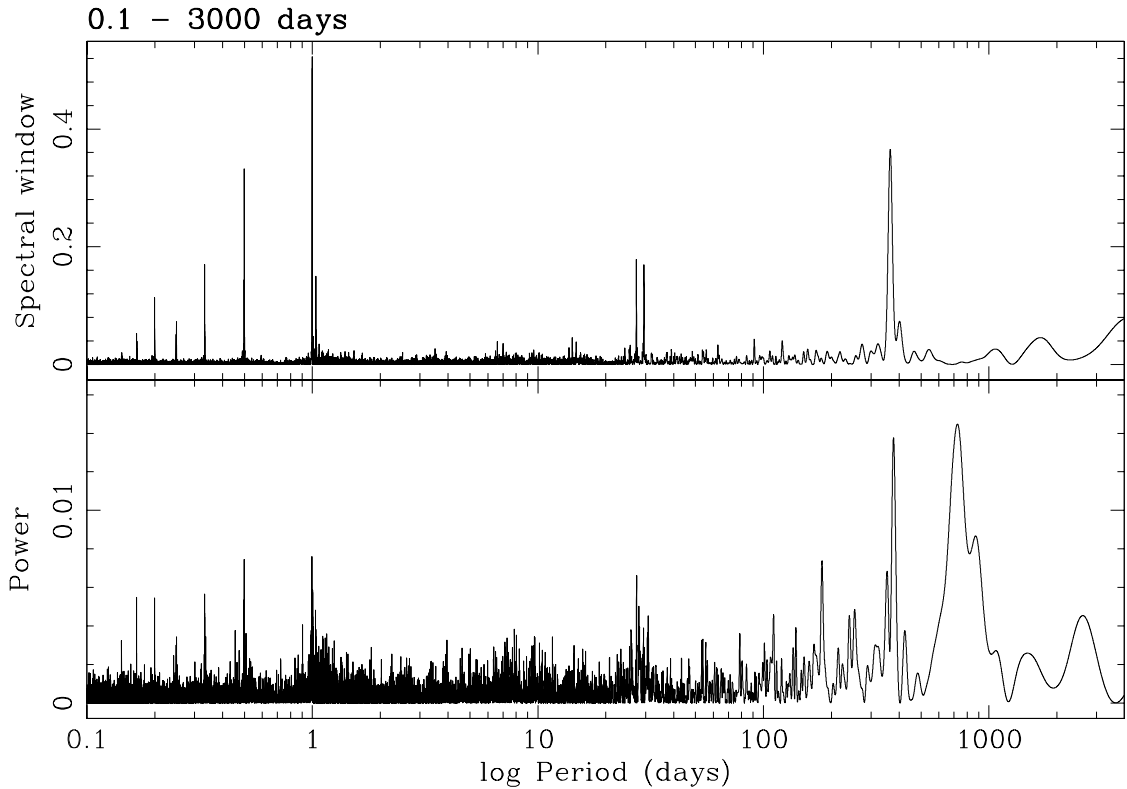}\\
  \end{center}
  \caption{Period determination following Fourier analysis and fine tuned based on visual inspection and $\mathcal{X}^2$ tests of the resulting light curves. Data windowing result ({\it top}) and power spectrum result ({\it bottom}), showing a clear peak at 737 days and the expected alias around 1, 27, 29 and 365 days (in the spectral window), together with their n-harmonics. The power spectrum is dominated by two strong periods, not due to interference from the spectral window at 376 and 737 days.}
  \label{fig:period}
\end{figure*}

To search for any periodic signals within the Harvard data we employed various Fourier analysis techniques using only those data with actual detections (not the upper limits).  Within the period range of $0.1-3000$ days, we visually inspected all folded light curves around peaks in the Fourier power spectrum (see Figure \ref{fig:period}).  To further fine-tune any significant periods $\mathcal{X}^2$ tests of the resulting light curves were performed.  Data windowing techniques were applied to flag any aliasing.  The data window simply depends upon the manner in which the sampling of the signal was performed and does not depend upon the signal itself.

\begin{figure}
\begin{center}
\includegraphics[angle=270,width=0.8\columnwidth]{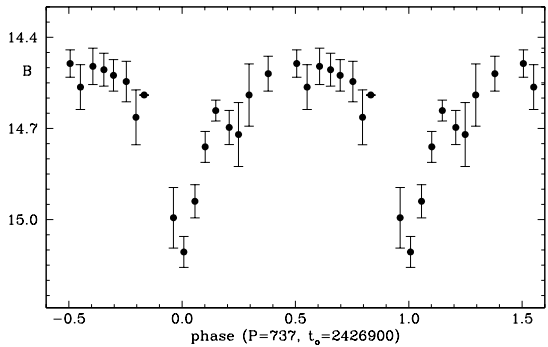}
\end{center}
  \caption{Photographic data folded onto the 737 days  (t$_0 = $ 2426900 HJD) periodicity. The data have been divided into 20 bins and for each bin an average phase and magnitude has been computed.}
  \label{fig:fold}
\end{figure}

We determined a primary period of 737 days (see Figure \ref{fig:period}) although a strong peak at 376 days is also observable, the latter produces a sinusoid-like light curve. In Figure~\ref{fig:fold} we present the photographic data folded at a period of 737 days (t$_0 = $ 2426900 HJD).  These data were divided into 20 bins and for each bin an average phase and magnitude were computed. The light curve in Figure~\ref{fig:fold} is reminiscent of both a reflection/heating effect \citep[e.g.][]{MJ02} and/or an eclipse.

The two above periods are outstanding in both the Fourier power spectrum and with all other period-searching programs tested. In addition we inspected by eye some tens of thousands of in-phase light curves plotted carefully exploring around these two periods, and found nothing additional. The 737 day period is particularly evident in the direct light curve (Figure~\ref{fig:fig1}).

In symbiotic binaries, a cool giant is in orbit with a massive white dwarf.  The distribution of their orbital periods peaks at around two years \citep{M03}, their cool giants usually fill the Roche lobe producing photometric modulations of half the orbital period \citep{MKS03}, and they frequently display strong [Ne~V]~3426\AA\ and [Ne~III]~3868\AA\ emission lines in quiescence \citep{MZ02}. The 737 and 376 day periods found in KT Eri are very close to the 2:1 ratio, which suggest by analogy with symbiotic stars, that its cool giant is filling its Roche lobe and the orbital inclination is large.

\subsection{Progenitor system}

A search through the NASA/IPAC IRSA allowed for the identification of three distinct sources at, or near, the location of KT~Eri, which may be three separate detections of the progenitor system: 2MASS J04475419-1010429, DENIS J044754.2-101042 and USNO B1 0798-0048707. The apparent magnitudes of the ``progenitor'' from each source are presented in Table~\ref{tb:progenitor}. We choose to use 2MASS infrared magnitudes because this spectral region is more sensitive to the secondary star. We then convert these to absolute magnitudes using the distance and reddening of $d=6.5$~kpc and $E_{B-V}=0.08$ \citep{RBS09} respectively. We thus derive $M_{J}=0.48\pm0.03$, $M_{H}=0.05\pm0.05$ and $M_{K_S}=0.00\pm0.07$. These magnitudes are brighter than those derived for the RN U~Sco at quiescence from 2MASS ($M_{J}=1.3\pm0.4$, $M_{H}=0.9\pm0.4$ and $M_{K_S}=0.9\pm0.4$, \citealt{DRB11}, see also Figure~\ref{fig:diagram}) and significantly brighter than typical quiescent CNe \citep{DRW11}. Furthermore, the colours are more in line with those of cool giant stars.

\begin{table}
  \caption{A summary of photometric observations of the KT~Eri progenitor system at quiescence.}
  \begin{center}
   \begin{tabular}{lllll}
    \hline\hline
    Telescope & Filter & Time (Days & Phase & Magnitude \\
              &        & before max) & & \\
    \hline
    Harvard & $B_{pg}$ & - & 0.20 & 14.68$\pm$0.40\tablefootmark{a} \\
    \hline
    USNO-B1& $B$ & 9,895 & 0.92 & 15.19$\pm$0.30 \\
    USNO-B1& $R$ & 6,932 & 0.94 & 15.31$\pm$0.30 \\
    USNO-B1& $I$ & 3,626 & 0.42 & 14.71$\pm$0.30 \\
    \hline
    2MASS & $J$ & 3,990 & 0.93 & 14.62$\pm$0.03 \\
    2MASS & $H$ & 3,990 & 0.93 & 14.16$\pm$0.05 \\
    2MASS & $K_S$ & 3,990 & 0.93 & 14.09$\pm$0.07 \\
    \hline
  \end{tabular}
  \end{center}
  \tablefoot{
  \tablefoottext{a}{Quoted error is one standard deviation.}
  }
\label{tb:progenitor}
\end{table}

\section{Discussion}\label{discussion}

The mean quiescent $B$-band luminosity determined from the Harvard plates, assuming a distance of 6.5~kpc and $E_{B-V}=0.08$, is $3.80\times10^{35}$~erg~s$^{-1}$. From this luminosity we can estimate a recurrence timescale, $\Delta T$, for the system, assuming that the $B$-band luminosity arises mainly from the accretion disc, and relating this to the critical mass for ignition and accretion rate, $M_{\mathrm{crit}}$ and $\dot{M}_{\mathrm{acc}}$ respectively ($\Delta T \sim M_{\mathrm{crit}} / \dot{M}_{\mathrm{acc}}$). This luminosity is similar to that derived for V2491~Cyg \citep{POE10} and using their Figure~8 we can estimate roughly the recurrence timescale for KT~Eri. We should note as a caveat that \citet{DRB11} indicate that the luminosity is much greater if the distance to V2491~Cyg is assumed to be 14~kpc rather than 10.5~kpc \citep[][respectively]{MSD11,HWV08}. The \citeauthor{POE10} figure provides us with a recurrence timescale ranging from a few decades for the highest $M_{\mathrm{WD}}\sim1.4 \mathrm{M}_{\odot}$ to $10^4$ years for $M_{\mathrm{WD}}\sim1 \mathrm{M}_{\odot}$. We can constrain $M_{\mathrm{WD}}<1.3 \mathrm{M}_{\odot}$, following the fact the time for the emergence of the SSS is much later than systems like RS~Oph and U~Sco \citep[][respectively, see Bode et al. 2011, in preparation]{HKL07,KHP99}, with known WD masses close to the Chandrasekhar limit, which would imply a $\Delta T >100$~years, depending on the mass of the WD and accretion rate. Such a long recurrence time may be consistent with the lack of a previous outbursts observed in the Harvard data (see Figures~\ref{fig:fig1} and \ref{fig:stats}).

\begin{figure}
\begin{center}
\includegraphics[width=0.8\columnwidth]{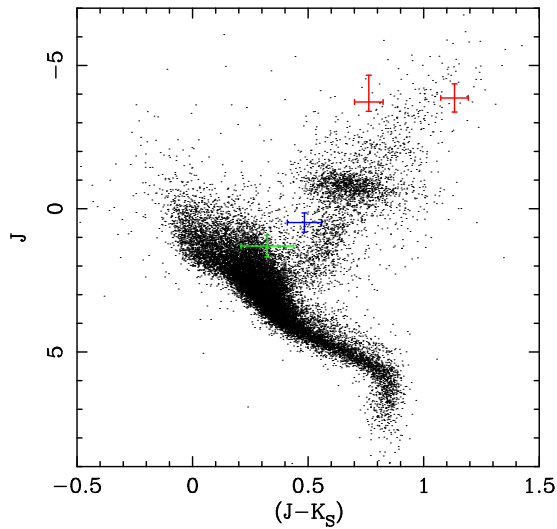}
\end{center}
\caption{Colour-magnitude diagram using {\it Hipparcos} data \citep{PE97}.  The green point indicates the position of a quiescent U~Sco, and the red point the location of a quiescent RS~Oph (left) and T~CrB (right).  The blue point shows the position of a quiescent KT~Eri.  Although the position of U~Sco appears coincident with the upper main sequence, the system contains a sub-giant secondary.}
  \label{fig:diagram}
\end{figure}

In Figure~\ref{fig:diagram} we show the $J-K_S$ versus $J$ colour-magnitude diagram populated with nearby stars from the {\it Hipparcos} catalogue \citep{PE97} which aids in determining the nature of the progenitor system \citep{DRW11}. The quiescent positions of the RNe harbouring a red-giant secondary RS~Oph and T~CrB (red) and U~Sco (green) which harbours a sub-giant, are included for comparative purposes to KT~Eri (blue). It is evident that the KT~Eri system lies in between these systems as further evidence for an evolved secondary. Furthermore, as mentioned above, the infrared colours are consistent with giant stars however, not of RS Oph or T CrB which harbour a M giant. Indeed the $I$-band magnitude is comparable to a Red Clump Star which generally are K giants. We should also note that the accretion disk will have some effect on the $I$-band luminosity.

From the evidence provided above it is hard to escape from the conclusion that the secondary star is evolved. \citet{R10} showed early spectra of the outburst (13 $-$ 73 days) which did not show evidence of deceleration, expected as the ejecta hit any pre-existing circumstellar wind. However, \citet{MJA11} showed that for V407 Cyg, whose secondary is a Mira blowing a thick wind, by the second week after outburst all signatures of the wind/ejecta interaction had disappeared.

It is noteworthy that \citet{NMR09} presented a very low resolution spectrum of the KT Eri progenitor on 1971 January 25 (phase 0.1) from the Digitized First Byurakan Survey \citep{MNR07}. A strong UV continuum and several emission lines were detected (most interesting are [Ne~V]~3426\AA\ and [Ne~III]~3868\AA) which is indicative of a hot star with circumstellar material \citep{NMR09}. These lines have also been observed in the post-outburst spectrum of some novae for example RS Oph. Therefore, if the interpretation and calibration of the low resolution spectrum is correct, it could suggest Ne enrichment, which is not produced in the outburst but dredged up from the underlying WD which constrains the mass of the WD to $\ge$ 1.1~M$_{\odot}$. The magnitude derived from the integrated spectrum of B~=~14.31 is not what we would expect for the phase (0.1, Figure~\ref{fig:fold}) of the spectrum, however this is still within the quiescent magnitude of the system and therefore we still maintain that this is a quiescent spectrum.

The mean magnitude from the Harvard plates $B=14.7$ mag and peak at outburst, from white light,  $m_{\mathrm{SMEI}}=5.42$, imply an outburst amplitude $\Delta m\sim9$ magnitudes, which is similar to the outburst amplitude observed in U~Sco \citep{MZT99}. In the RS~Oph-class RNe their typical outburst amplitude is $\sim5-7$ magnitudes, due to the presence of a red-giant secondary, and in CNe up to $\Delta m\sim17$ magnitudes for the very fastest \citep{W08}.

\section{Conclusions}\label{conclusions}
In recent years the study of novae, particularly the recurrents, has intensified, beginning with the 2006 eruption of RS~Oph \citep{2008ASPC..401.....E}, the predicted 2010 eruption of U~Sco \citep{2010AJ....140..925S} and the discovery of new candidate RN systems \citep[for a comprehensive review see][]{S10}.  The nature of nova V2487~Oph, which included pre-outburst X-ray emission, indicated that this system may be a RN although only one outburst had been observed \citep{HKK02,HS02}.  The recurrent nature of V2487~Oph was confirmed with the identification of a previous outburst on 1900 June 20 by \citet{PSX09} who also predicted a typical recurrence timescale of 18 years (although only two outbursts had been observed separated by 118 years).  Furthermore, \citet{POE10} suggested that V2491~Cyg may recur on a timescale of $\sim$ 100 years whilst \citet{DRB11} show that the system is similar to U~Sco and may have a recurrence timescale of $<100$ years.   Such observations imply that the underlying nova population may exhibit a continuous range of recurrence times; rather than simply being grouped into very short timescale systems (RNe) and very long timescales (CNe).

We summarise our main results as follows:
\begin{itemize}
\item We found no previous outburst of KT~Eri between 1883 and 1952 but suggest that if KT~Eri is a RN it will have a recurrence timescale $\Delta T>100$ years. Nonetheless, we recommend archival plate searches around the 1970$-$80's in order to look for any previous outburst with  $\Delta T \approx 40$ years, similar to RN LMC~2009a (Bode et al., in preparation).
\item We find a periodicity of 737 days at quiescence. Combining it with the position of the KT Eri progenitor on the IR colour-magnitude diagram (Figure~\ref{fig:diagram}), the donor star appears to be a RGB star, although not as luminous (or evolved) as RS~Oph/T~CrB.  However, the possibility that the secondary is a red clump star cannot be specifically excluded by this study.
\item An outburst amplitude of $\sim9$ magnitudes is more akin to a the RN class of objects than a very fast CN, for which a $\sim17$ magnitude amplitude is expected, and thus may provide further evidence of the nature of the progenitor system as an evolved secondary.
\end{itemize}

We urge continued observations of this interesting object and further exploration of archival plates for any evidence of a previous outburst and/or interesting variability and in particular identifying the nature of the secondary star.

\begin{acknowledgements}
We would like to thank Alison Doane, Curator of Astronomical Photographs at the Harvard College Observatory, for use of their facilities and Andy Newsam for useful discussion. We thank an anonymous referee for valuable comments on the original manuscript. This research has made use of the NASA/IPAC Infrared Science Archive, which is operated by the Jet Propulsion Laboratory, California Institute of Technology, under contract with the National Aeronautics and Space Administration. The DENIS project has been partly funded by the SCIENCE and the HCM plans of the European Commission under grants CT920791 and CT940627. It is supported by INSU, MEN and CNRS in France, by the State of Baden-W\"urttemberg in Germany, by DGICYT in Spain, by CNR in Italy, by FFwFBWF in Austria, by FAPESP in Brazil, by OTKA grants F-4239 and F-013990 in Hungary, and by the ESO C\&EE grant A-04-046. Jean Claude Renault from IAP was the Project manager. Observations were carried out thanks to the contribution of numerous students and young scientists from all involved institutes, under the supervision of  P. Fouqu\'e, survey astronomer resident in Chile.  VARMR was supported by an STFC PhD studentship and is supported by a South African SKA Fellowship.

\end{acknowledgements}

\balance


\begin{thebibliography}{}

\bibitem[Allen(1984)]{A84} Allen, D.~A.\ 1984, Proceedings 
of the Astronomical Society of Australia, 5, 369 

\bibitem[Anupama(2008)]{A08} Anupama, G.~C.\ 2008, in ASP Conf. Ser., 401 eds. A. Evans, M. F. Bode, T. J. OÕBrien, \& M. J. Darnley, 31

\bibitem[Bode(2010)]{B10} Bode, M.~F.\ 2010, Astronomische 
Nachrichten, 331, 160 

\bibitem[Bode \& Evans(2008)]{BE08} Bode, M.~F., \& Evans, A.\ 2008, Classical Novae, Cambridge Astrophysics Series, Vol~43, Cambridge University Press, Cambridge

\bibitem[Bode et al.(2009)]{BOP09} Bode, M.~F., et al.\ 2009, 
The Astronomer's Telegram, 2025 

\bibitem[Bode et al.(2010)]{BOP10} Bode, M.~F., et al.\ 2010, 
The Astronomer's Telegram, 2392

\bibitem[Darnley et 
al.(2011a)]{DRB11} Darnley, M.~J., Ribeiro, V.~A.~R.~M., Bode, M.~F., \& Munari, U.\ 2011, \aap, 530, A70 

\bibitem[Darnley et al.(2011b)]{DRW11} Darnley, M.~J., Ribeiro, V.~A.~R.~M., Williams, R.~P., Hounsell, R. \& Bode, M.~F.\ 2011, \apj, submitted 

\bibitem[Drake et al.(2009)]{DDG09} Drake, A.~J., et al.\ 2009, The Astronomer's Telegram, 2331

\bibitem[\protect\citeauthoryear{Evans et al.}{2008}]{2008ASPC..401.....E} 
Evans, A., Bode, M.~F., O'Brien, T.~J., Darnley, M.~J.\ 2008, ASP Conf. Ser., 401

\bibitem[Eyles et al.(2003)]{ESC03} Eyles, C.~J., et al.\ 2003, Solar Phys., 217, 319

\bibitem[Guido et al.(2009)]{GSM09} Guido, E., Sostero, G.,  Maehara, H., \& Fujii, M.\ 2009, Central Bureau Electronic Telegrams, 2053

\bibitem[Hachisu et al.(2007)]{HKL07} Hachisu, I., Kato, M., 
\& Luna, G.~J.~M.\ 2007, \apjl, 659, L153 

\bibitem[Hachisu et al.(2002)]{HKK02} Hachisu, I., Kato, M., 
Kato, T., 
\& Matsumoto, K.\ 2002, The Physics of Cataclysmic Variables and Related Objects, 261, 629 

\bibitem[Helton et al.(2008)]{HWV08} Helton, L.~A., Woodward, C.~E., Vanlandingham, K., \& Schwarz, G.~J.\ 2008, Central Bureau Electronic Telegrams, 1379

\bibitem[Hernanz 
\& Sala(2002)]{HS02} Hernanz, M., \& Sala, G.\ 2002, Science, 298, 393 

\bibitem[Hounsell et al.(2010)]{HBH10} Hounsell, R., et al.\ 2010, \apj, 724, 480 

\bibitem[Itagaki(2009)]{I09} Itagaki, K.\ 2009, Central Bureau Electronic Telegrams, 2050

\bibitem[Jurdana-{\v S}epi{\'c} \& Munari(2010)]{JM10} Jurdana-{\v S}epi{\'c}, R., \& Munari, U.\ 2010, \pasp, 122, 35 

\bibitem[Kahabka et 
al.(1999)]{KHP99} Kahabka, P., Hartmann, H.~W., Parmar, A.~N., \& Negueruela, I.\ 1999, \aap, 347, L43 

\bibitem[Maehara(2009)]{M09} Maehara, H.\ 2009, Central Bureau Electronic Telegrams, 2053

\bibitem[Mickaelian et al.(2007)]{MNR07} Mickaelian, A.~M., et al.\ 2007, \aap, 464, 1177 

\bibitem[Miko{\l}ajewska(2003)]{M03} Miko{\l}ajewska, J.\ 
2003, in ASP Conf. Ser., 303, eds. R. L. M. Corradi, R. Miko{\l}ajewska \& T. J. Mahoney, 9 

\bibitem[Miko{\l}ajewska et al.(2003)]{MKS03} 
Miko{\l}ajewska, J., Kolotilov, E.~A., Shugarov, S.~Y., Tatarnikova, A.~A., 
\& Yudin, B.~F.\ 2003, in ASP Conf. Ser., 303, eds. R. L. M. Corradi, R. Miko{\l}ajewska \& T. J. Mahoney, 151 

\bibitem[Monet et al.(2003)]{MLC03} Monet, D.~G., et al.\ 
2003, \aj, 125, 984 

\bibitem[Munari et 
al.(1999)]{MZT99} Munari, U., et al.\ 1999, \aap, 347, L39 

\bibitem[Munari et al.(2011a)]{MSF11} Munari, U., et al.\ 2011a, in ``Asiago Meeting on Symbiotic Stars'', Baltic Astronomy special issue, in press 

\bibitem[Munari et al.(2011b)]{MJA11} Munari, U., et al.\ 2011b, \mnras, 410, L52

\bibitem[Munari et al.(2001)]{MJM01} Munari, U., Jurdana-{\v S}epi{\'c}, R., \& Moro, D.\ 2001, \aap, 370, 503 

\bibitem[Munari \& Jurdana-{\v S}epi{\'c}(2002)]{MJ02} Munari, U., \& Jurdana-{\v S}epi{\'c}, R.\ 2002, \aap, 386, 237

\bibitem[Munari et al.(2011c)]{MRB11} Munari, U., Ribeiro, V. A. R. M., Bode, M. F., \& Saguner, T.\ 2011c, \mnras, 410, 525

\bibitem[Munari et al.(2011d)]{MSD11} Munari, U., Siviero, A., Dallaporta, S., Cherini, G., Valisa, P., \& Tomasella, L.\ 2011d, \na, 16, 209 

\bibitem[Munari \& Zwitter(2002)]{MZ02} Munari, U., \& Zwitter, T.\ 2002, \aap, 383, 188 

\bibitem[Nesci et al.(2009)]{NMR09} Nesci, R., Mickaelian, 
A., \& Rossi, C.\ 2009, The Astronomer's Telegram, 2338

\bibitem[O'Brien et al.(2010)]{OMS10} O'Brien, T.~J., Muxlow,  T.~W.~B., Stevens, J., Datta, A., Roy, N., Eyres, S.~P.~S., \& Bode, M.~F.\ 2010, The Astronomer's Telegram, 2434 

\bibitem[Page et al.(2010)]{POE10} Page, K.~L., et al.\ 2010, 
\mnras, 401, 121 

\bibitem[Pagnotta et al.(2009)]{PSX09} Pagnotta, A., 
Schaefer, B.~E., Xiao, L., Collazzi, A.~C., 
\& Kroll, P.\ 2009, \aj, 138, 1230 

\bibitem[Perryman \& ESA(1997)]{PE97} Perryman, M.~A.~C., \& ESA 1997, ESA Special Publication, 1200

\bibitem[Ragan et al.(2009)]{RBS09} Ragan, E., et al.\ 2009, The Astronomer's Telegram, 2327

\bibitem[Ribeiro(2011)]{R10} Ribeiro, V. A. R. M.\ 2011, Ph.D.~Thesis, Liverpool John Moores University

\bibitem[Ribeiro et al.(2011)]{RDB11} Ribeiro, V. A. R. M., Darnley, M. J., Bode, M. F., Munari, U., Harman, D. J., Steele, I. A., \& Meaburn, J.\ 2011, \mnras, 412, 1701

\bibitem[Ribeiro et al.(2009)]{RBD09} Ribeiro, V.~A.~R.~M., 
et al.\ 2009, \apj, 703, 1955 

\bibitem[Rudy et al.(2009)]{RPR09} Rudy, R.~J., Prater,  T.~R., Russell, R.~W., Puetter, R.~C., \& Perry, R.~B.\ 2009, Central Bureau Electronic Telegrams, 2055

\bibitem[Schaefer(2010)]{S10} Schaefer, B.~E.\ 2010, \apjs, 
187, 275 

\bibitem[\protect\citeauthoryear{Schaefer et 
al.}{2010}]{2010AJ....140..925S} Schaefer, B.~E., et al. 2010, AJ, 140, 925 

\bibitem[Skrutskie et al.(2006)]{SCS06} Skrutskie, M.~F., et 
al.\ 2006, \aj, 131, 1163 

\bibitem[Slavin et al.(1995)]{SOD95} Slavin, A.~J., O'Brien, 
T.~J., \& Dunlop, J.~S.\ 1995, \mnras, 276, 353 

\bibitem[Warner(2008)]{W08} Warner, B. 2008, in Classical Novae, eds. M. F. Bode \& A. Evans, Cambridge Astrophysics Series, Vol 43, Cambridge University Press, 16

\bibitem[\protect\citeauthoryear{Yaron et al.}{2005}]{2005ApJ...623..398Y} 
Yaron O., Prialnik D., Shara M.~M., Kovetz A., 2005, ApJ, 623, 398 


\end{thebibliography}
\end{document}